\documentclass[12pt]{article}
\usepackage{amsmath,amssymb}
\usepackage{graphicx}
\usepackage{color}
\allowdisplaybreaks[1]

\makeatletter
 
  \@addtoreset{equation}{section}
 \makeatother

\newcommand{\R}{\mathbb{R}}

\newcommand{\C}{\mathbb{C}}

\newcommand{\nn}{\nonumber}

\newcommand{\Tr}{{\rm Tr}\,}

\newcommand{\del}{\partial}

\newcommand{\hsigma}{\hat{\sigma}}

\newcommand{\bv}{{\mathbf v}}
\newcommand{\bw}{{\mathbf w}}
\newcommand{\bX}{{\mathbf X}}
\newcommand{\bA}{{\mathbf A}}

\newcommand{\bM}{{\mathbf M}}

\def\bea#1\ena{\begin{align}#1\end{align}}
\newcommand{\Bra}[1]{\left\langle\, #1\,\right|}
\newcommand{\Ket}[1]{\left|\, #1\,\right\rangle}
\newcommand{\Bracket}[2]{\left\langle\, #1\,|\, #2\,\right\rangle}
\def\mat#1{\matt[#1]}
\def\matt[#1,#2,#3,#4]{\left(%
\begin{array}{cc} #1 & #2 \\ #3 & #4 \end{array} \right)}
\def\vect#1{\vectt[#1]}
\def\vectt[#1,#2]{\left(%
\begin{array}{c} #1 \\ #2 \end{array} \right)}

\def\bc#1{\textcolor{blue}{#1}}

\allowdisplaybreaks
\begin{document}
\begin{titlepage}

\setcounter{page}{0}
\renewcommand{\thefootnote}{\fnsymbol{footnote}}

\begin{center}
{\large\bf
Spherical D-brane by Tachyon Condensation
}

\vspace{10mm}
Tsuguhiko Asakawa$^{1}$%
\footnote{\tt asakawa@maebashi-it.ac.jp}
and So Matsuura$^{2}$%
\footnote{\tt s.matsu@phys-h.keio.ac.jp}
\vspace{10mm}

{\em
$^{1}$ Department of Integrated Design Engineering,\\
Maebashi Institute of Technology,
Maebashi, 371-0816, Japan \\
$^{2}$ Department of Physics, Hiyoshi Campus, 
and Research and Education Center for Natural Science, \\
Keio University, 4-1-1 Hiyoshi, Yokohama, 223-8521, Japan\\
}
\end{center}

\vspace{20mm}
\centerline{{\bf Abstract}}
\vspace{3mm}
{\small
We {find} a novel tachyon condensation 
which provides a D-brane system with spherical worldvolume
in the flat spacetime.
The tachyon profile is a deformation of a known D0-brane solution
on non-BPS D3-branes in type IIA superstring theory, 
which {realizes} a bound state of a spherical D2-brane and a D0-brane. 
{The} D0-brane is resolved into the sphere as a $U(1)$ monopole flux of
the unit magnetic charge.
We {show} that the system has the correct tension and the RR-coupling. 
{Although the low energy effective action of the system is the same as that of
the} dual description of the fuzzy sphere solution of multiple D0-branes,
our system cannot be equivalent to the fuzzy sphere.
We {argue} that the sphere can be stable in a certain RR-flux background.
The use of projective modules in describing the tachyon condensation is emphasized.}
\end{titlepage}
\newpage

\setcounter{footnote}{0}

\section{Introduction}
\label{sec:introduction}

The spectrum of superstring theory includes not only the BPS D-branes
but also unstable D-brane systems
such as non-BPS D-branes and D-brane-anti-D-brane systems of various dimensions.
Their instability is controlled by the presence of tachyon fields on them.
The system decays into the vacuum when the tachyon field acquires
a vacuum expectation value on the whole unstable D-brane system.
If the tachyon has a solitonic profile, on the other hand,
the system reduces to lower dimensional BPS D-brane(s)
(D-brane descent relation) \cite{Sen1998},
which leads to the beautiful theory of the classification of the RR-charges
for BPS D-branes by K-theory \cite{Witten1998,Horava1998}.%
\footnote{
There is also an opposite direction (D-brane ascent relation)
where a higher dimensional D-brane are constructed by lower dimensional
unstable D-brane systems \cite{Asakawa2002}.
In this setting, compact worldvolumes are treated rather easier
because a D-brane corresponds to a spectral triple
in noncommutative geometry \cite{Connes1995}.
See \cite{Terashima2005,Ellwood2005}
for a spherical D-brane in this setting.}
Although it is sufficient for the classification problem
to consider the representatives, 
such as the Atiyah-Bott-Shapiro (ABS) construction \cite{Atiyah1964},
there are indeed various possible profiles which carry the same RR-charges.
In this paper, we study the tachyon condensation of unstable D-branes which
descends to a D-brane with spherical worldvolume (spherical D-brane)
in the Minkowski spacetime
as an example of a compact worldvolume in non-compact spacetime.
There is no such construction before in the literature as far as the authors know.

The dielectric D-brane \cite{Myers1999} is known as in
a class of spherical D-branes.
In a certain RR-background, the effective theory on multiple D0-branes
is accompanied with the so-called Myers term,
which makes it possible to exist a stable fuzzy sphere solution.
It is interpreted as a noncommutative D2-brane
and is called dielectric because it couples to the RR-field through
its field strength.
This construction requires at least two D0-branes (${k}=2$) and
it reduces to a commutative {sphere} with ${k}$ magnetic flux
by taking an appropriate continuum (large ${k}$) limit.
Our construction is somehow complementary to this construction,
that is, we obtain
a commutative sphere with the unit (${k}=1$) magnetic flux
without taking any continuum limit.

To be specific, we consider an effective theory on two non-BPS D$3$-branes
with spatial worldvolume ${\mathbb R}^3$
and give a tachyon profile corresponding to a spherical D$2$-brane.
Our profile is just a constant shift of the known single D$0$-brane solution.
It is remarkable that
this shift changes the spatial worldvolume of the system
from a point to $S^2$.
We see that the profile behaves more like a kink in the radial direction
different from the D$0$-brane as a soliton.
On $S^2$, there is also a unit magnetic flux carrying a D$0$-brane charge,
which is the same $U(1)$ flux as a Wu-Yang monopole \cite{Wu1976}
sitting at the origin in ${\mathbb R}^3$.
Thus the resulting system is in fact a D$2$-D$0$ bound state.

The tachyon condensation for BPS bound states
is studied already in \cite{Hashimoto2005a} and \cite{Hashimoto2006},
where the Nahm construction of monopoles \cite{Nahm1980}
and the ADHM construction of gauge instantons \cite{Atiyah1978}, respectively,
are realized by tachyon condensation in unstable D-brane systems.
Our analysis is in the same spirit with these works.
We show that
the effective action of the spherical D$2$-brane is reproduced
by inserting the tachyon profile into
the effective theory of non-BPS D$3$-branes,
together with the correct tension and the RR-coupling for a D2-D0 bound state.
An analysis on the stability of the spherical D2-brane is also given
by using the method developed in 
\cite{Myers1999,Hashimoto2002,Hyakutake2005}.
As shown in these papers,
the size of the sphere can be stable in the constant RR-flux background
but in a delicate way.
In our case this size of the sphere is in the string scale 
in a constant flux background.
However, we will show that the sphere can have a stable macroscopic
radius by considering another RR-background of radially graduating flux.

The organization of the paper is as follows.
In the section 2,
we review the construction of a BPS D0-branes from non-BPS D3-branes
by the tachyon condensation.
Then, we construct a spherical D2-brane with a monopole flux
by deforming the tachyon profile of the single D0-brane solution.
The mechanism to appear the non-trivial gauge potential
can be clearly understood
by using the language of the projective module.
In the section 3, we
obtain the effective action of the spherical D2-brane
from that of non-BPS D3-branes,
which shows that the system indeed possesses
the correct RR-charges and tension of a D2-D0 bound state.
We examine the stability of the sphere in two different RR-flux backgrounds.
The section 4 is devoted to the conclusion and the discussion.
In the appendix \ref{app:radial delta},
we summarize the properties of radial delta function.
In the appendix \ref{app:BSFT},
we summarize the construction of the spherical D2-brane
based on the boundary string field theory (BSFT).

\section{Spherical D-brane via tachyon condensation}
\label{sec:spherical brane}

\subsection{Non-BPS D3-branes and D0-brane solution}
\label{sec:setting}

Let us start with recalling the construction of a BPS D0-brane
in order to fix our setup and notations.
We consider
two non-BPS $D3$-branes in Type IIA theory.
The worldvolume of the non-BPS D3-branes
is $\R^{4}$ and the low energy effective theory is a
$U(2)$ gauge theory
with gauge fields, transverse scalar fields, 
a real tachyon field and fermions
which belong to the adjoint representation of $U(2)$.
Throughout of this paper, we do not consider the scalar fields and the fermions
for simplicity. 
In addition, we do not consider the gauge field in the first setup and 
focus only on static configurations for the tachyon field.
With this restriction, the relevant field is
a tachyon field $T(x)$ which is a $2\times 2$ hermitian matrix-valued function
on the spatial worldvolume $\R^3$.
Note also that the Chan-Paton bundle $\R^3 \times \C^2$ is trivial.

For the effective action of the non-BPS D3-branes, we adopt the BSFT action
which has the form,
\begin{align}
&S[T]=S_{\rm NSNS}[T]+S_{\rm RR}[T],
\label{D3 action}
\end{align}
with roughly
\begin{align}
\begin{split}
&S_{\rm NSNS}[T]=\sqrt{2}T_3 \int_{{\R}^4} d^4x ~{\rm Tr}_2 \left(e^{-T^2}{\cal F}(\partial T)\right),\\
&S_{\rm RR}[T]=\sqrt{2}\mu_3 \int_{{\R}^4} C\wedge {\rm Str}_2 \left(e^{-T^2+dT}\right),
\end{split}
\end{align}
where each term is obtained by the disc amplitude in the NSNS- and RR-sector, respectively,
$T_3=\mu_3$ is the tension for a BPS D3-brane,
${\cal F}(\partial T)$ is some polynomial in the derivative of $T$
and
${\rm Str}_2$ stands for the trace but is taken only for odd wedge powers of $dT$.
The detail of this action is summarized in the Appendix \ref{app:BSFT}.
The following argument can be applied to any other effective actions obtained by field redefinitions of $T$.

Here we focus on the tachyon potential $e^{-T^2}$
in order to capture the essence of our construction%
\footnote{The potential is usually $\Tr_2 e^{-T^2}$
but we use the same terminology for the expression without the trace.}.
It shows the instability of the system of non-BPS D3-branes.
At the true vacuum $T(x)=u \,{\rm diag}(\pm 1,\pm 1)$ ($u\to\infty$),
the system annihilates to the closed string vacuum \cite{Sen1998}.

Next, recall that
the tachyon profile of the codimension three soliton,
\begin{equation}
T(x)=u\sigma_i x^i
\label{D0 profile}
\end{equation}
with the Pauli matrices $\sigma_i$ $(i=1,2,3)$
represents a BPS D0-brane sitting at the origin of the $\R^3$
in the limit of $u\to\infty$,
which is know as the ABS construction \cite{Atiyah1964}.
This is roughly seen by the tachyon potential,
\begin{align}
e^{-T^2}
=e^{-u^2|x|^2} 1_2
\xrightarrow{u\to\infty} \left(\frac{\sqrt{\pi}}{u}\right)^3\delta^{(3)}(x) 1_2,
\label{potential with D0}
\end{align}
where $1_2$ denotes the unit matrix%
\footnote{In the following, we sometimes omit $1_2$ for notational simplicity.}
and
$\delta^{(3)}(x)=\lim_{u\to\infty}(\frac{u}{\sqrt{\pi}})^3 e^{-u^2|x|^2}$
is used.
The appearance of the {Dirac} delta function {on $\R^3$}
shows that the remnant after the condensation is left at the origin,
which corresponds to the zero of $T^2$ {(i.e., false vacuum)}.
In general, condensation defects
correspond to the zeros of the tachyon profile.

This profile is indeed a classical solution of the effective action
in the $u\to\infty$ limit, 
which reproduces the correct tension and RR-charge for a BPS D0-brane
when inserting \eqref{D0 profile} to the action \eqref{D3 action}
\cite{Kraus2001,Takayanagi2000}.
The $u$-dependent factor in \eqref{potential with D0} is important
to obtain the tension.
On the other hand, due to its topological nature,
the Chern-Simons (CS) term is independent of $u$.

\subsection{Construction of spherical D2-brane}
\label{sec:construction}

In this paper, we propose that a single spherical
D2-brane is given by shifting
the D0-brane solution (\ref{D0 profile}) by a constant $R>0$ as
\begin{align}
T(x)=u \left(\sigma_i x^i -R \right).
\label{profile}
\end{align}
Although the modification is quite simple, it causes two effects:
the profile represents a two dimensional spatial worldvolume $S^2$
with radius $R$ and a monopole gauge flux appears on the $S^2$,
as we will show.
Note that this deformation is {\it not} regarded as a fluctuation around
the D0-brane solution \eqref{D0 profile}.
In fact, the fluctuation corresponding to scalar fields $\Phi^i$ on a D0-brane
is incorporated as $T=u \sigma_i (x^i-\Phi^i)$.

In order to see how
a spherical object appears by the tachyon condensation,
it is worth looking at
the tachyon potential $e^{-T^2}$ closely.
To this end, it is convenient to diagonalize the
square of the profile \eqref{profile},
\begin{align}
T^2=u^2 \left( |x|^2+R^2-2R \sigma_i x^i\right), 
\end{align}
where $|x|\equiv \sqrt{x_1^2+x_2^2+x_3^2}$. 
Since the eigenvalues of $T^2$ are $u^2(|x| \pm R)^2$,
the tachyon potential can be transformed by
a suitable $SU(2)$ gauge transformation $U(x)$ as
\begin{align}
U^\dagger (x) e^{-T^2}U(x) = \mat{e^{- u^2(|x| - R)^2},0,0,e^{- u^2(|x| + R)^2}}
\xrightarrow{u\to\infty}\mat{\frac{\sqrt{\pi}}{u} \delta (r-R),0,0,0},
\end{align}
where $\delta (r-R)$ with $r=|x|$ denotes the radial delta function
{defined by }
\begin{align}
\lim_{u\to\infty}\frac{u}{\sqrt{\pi}}e^{- u^2(r - R)^2}=\delta (r-R).
\end{align}
{The detail of the radial delta-function is summarized 
in Appendix \ref{app:radial delta}.}
Since the radial delta function has support at $r=R$,
we observe that a non-BPS D3-brane corresponding to a upper component
produces a spherical object,
while the lower component disappears after the tachyon condensation.
Let us make this {rough sketch} more concrete in the following.

\subsubsection{Diagonalizing the tachyon potential}
Next let us specify the gauge transformation that diagonalizes $T^2$ at each
point $x$ in $\R^3$.
Instead of $T^2$, it is easier to find the two eigenvectors ${\bf v}$ and ${\bf w}$ of $T$ itself
corresponding to eigenvalues $u(|x|- R)$ and $-u(|x|+R)$, respectively.

{Such} vectors cannot be defined on the whole $\R^3$ but
exist only locally.
More precisely, we should consider two regions in $\R^3$:
${\cal U}_N=\R^3\backslash \{r=-x_3\}$ and
${\cal U}_S=\R^3\backslash \{r=x_3\}$.
At each point $x(\ne 0)  \in \R^3$,
the tachyon profile (\ref{profile}) can be diagonalized as
\begin{equation}
U_{N/S}(x)^{\dagger} T(x) U_{N/S}(x) =
u \left(\begin{matrix} r-R & 0 \\ 0 & -r-R  \end{matrix}\right),
\label{diagonal T}
\end{equation}
by using either of
\begin{align}
\begin{split}
U_{N}(x) \equiv (\bv_N,\bw_N)
= \frac{1}{\sqrt{ 2r(r+x_3)}}
\left(\begin{matrix}
r+x_3 & -\bar{z} \\ z & r+x_3
\end{matrix}\right)  & {~\text{for}~ x\in {\cal U}_N}, \\
U_{S}(x) \equiv (\bv_S,\bw_S)
= \frac{1}{\sqrt{ 2r(r-x_3)}}
\left(\begin{matrix}
\bar{z} & -r+x_3 \\
r-x_3 &  z
\end{matrix}\right) & {~\text{for}~ x\in {\cal U}_S},
\end{split}
\label{unitary matrix}
\end{align}
where 
$z=x_1+ix_2$ and $\bar{z}=x_1-ix_2$.
For each suffix $N$ and $S$,
$\bv$ and $\bw$ are orthonormal,
$\bv^\dagger \bv=1=\bw^\dagger \bw$, $\bw^\dagger \bv=0$,
and complete, $\bv\bv^\dagger +\bw\bw^\dagger =1_2$.
Since $\det U_{N/S}=1$, $U_{N}(x)$($U_S(x))$ is an element of $SU(2)$  at each point $x$.
Note that $U_{N}(x)$ ($U_{S}(x)$) is well-defined in ${\cal U}_N$ (${\cal U}_S$)
so that we need a patch-wise diagonalization, but the r.h.s. of \eqref{diagonal T}
is globally well-defined on whole $\R^3$ including the origin%
\footnote{Note that ${\cal U}_N \cup {\cal U}_S=\R^3\backslash \{r=0\}$
excludes the origin $r=0$, {but \eqref{profile} is diagonal there.}}.
We point out that $U_{N/S}$ are independent of the both parameters $u$ and $R$.

In the overlapping region ${\cal U}_N \cap  {\cal U}_S$ (other than {the} $x^3$-axis),
these two unitary matrices are related by the transition function,
\begin{align}
U_{N}^\dagger U_{S} &= \vect{\bv_N^\dagger, \bw_N^\dagger} (\bv_S,\bw_S)
=\frac{1}{|z|}\mat{\bar{z},0,0,z},
\label{diagonal U(1)}
\end{align}
which lies in a subgroup $U(1)$ of $SU(2)$.\\

The tachyon potential is now written as
\begin{align}
e^{-T^2}
=U_{N/S}(x) \mat{e^{- u^2(r- R)^2},0,0,e^{- u^2(r + R)^2}}U_{N/S}^\dagger (x)
\xrightarrow{u\to\infty}\frac{\sqrt{\pi}}{u} \delta (r-R)P_{N/S}(x),
\label{str of potential}
\end{align}
Here $P_{N}$ and $P_S$ are defined by
\begin{align}
P_{N/S}(x) =U_{N/S}(x) P_0 U_{N/S}^\dagger (x)  \quad\text{with}\quad P_0=\mat{1,0,0,0}.
\label{P}
\end{align}
The operators $P_{N/S}(x)$ are projection operators acting on the Chan-Paton space at each point $x$,
which are unitary equivalent to the matrix $P_0$.
This shows the structure of the tachyon condensation: 
The matrix $P_0$ represents that one of {the} 
D3-branes {annihilates} into the vacuum
and the delta function causes the localization of the spatial worldvolume of
another D3-brane from $\R^3$ to $S^2$.
As we will see soon below,
the gauge transformations $U_{N/S}$, or more precisely $P_{N/S}$,
induce the $U(1)$ monopole gauge flux on $S^2$.

\subsubsection{Monopole flux}

Although the appearance of a non-trivial gauge flux after
a gauge transformation sounds odd in ordinary gauge theory,
it happens in tachyonic theory.
The mechanism in our present setting is as follows \cite{Asakawa2006}:
Since the unitary transformation by $U_{N}(x)$
is a local gauge transformation,
it acts not only on the tachyon field but also on the $U(2)$
gauge field on non-BPS D3-branes.
Since there is no gauge field in the beginning, there appears the pure gauge
$-iU_{N}^\dagger dU_{N}$ on $\R^3$.
Note that there is no $u$ dependence in the gauge field.
Here it is important that
the tachyon potential (after diagonalization) with the form
$\delta (r-R)P_0$ acts on this pure gauge potential.
Since the projection operator $P_0$ picks up the upper-left component
$A_N=-i\bv_N^\dagger d\bv_N$ from $-iU_{N}^\dagger dU_{N}$
and $\delta (r-R)$ restricts the domain to ${\cal U}_N \cap S^2$,
a non-trivial gauge potential remains after 
taking the limit of $u\to\infty$.
The case of $U_S$ is similar.

As a result of this mechanism, the tachyon condensation produces
a non-trivial $U(1)$ gauge potential,
\begin{align}
  \begin{split}
& A_N = -i \bv_N^\dagger d \bv_N
= \frac{x^1 dx^2 - x^2 dx^1}{2r(r+x^3)} =\frac{1}{2}(1-\cos\theta )d\phi,\\
& A_S =-i \bv_S^\dagger d \bv_S
= \frac{x^2 dx^1 - x^1 dx^2 }{2r(r-x^3)}=-\frac{1}{2}(1+\cos\theta )d\phi,
\end{split}
\label{monopole potential}
\end{align}
on the open set ${\cal U}_N$ and ${\cal U}_S$,
respectively, where
we {have} used the polar coordinates in $\R^3$ in the last expressions.
They are the same as the Wu-Yang magnetic monopole of the unit magnetic charge
sitting at the origin of $\R^3$
where the excluded region in each open set ${\cal U}_{N/S}$
is nothing but the Dirac-string singularity.
They are related with each other by the $U(1)$ transition function $e^{-i\phi}$
(upper-left part of \eqref{diagonal U(1)})
in the overlapping region ${\cal U}_N \cap {\cal U}_S$.
The corresponding $U(1)$ field strength,
\begin{equation}
F=\frac{1}{2}\sin\theta d\theta \wedge d\phi,
\label{monopole flux}
\end{equation}
is globally defined on $\R^3\backslash \{0\}$.
In our case, because of the delta function,
\eqref{monopole potential} and \eqref{monopole flux} are restricted on $S^2$
and the flux is a uniform magnetic flux on $S^2$,
which represents the unit D0-brane on the spherical D2-brane
according to the {\it brane within brane} mechanism \cite{Douglas1995}.
We thus expect that the system after the tachyon condensation
is a bound state of a spherical D2-brane and a resolved D0-brane on it.
We will confirm this by evaluating the tension and the RR couplings of this system in the proceeding section.

\subsubsection{Projective module}
We can describe the above mechanism by using
the full use of BSFT
or boundary states \cite{Asakawa2006}, 
which is provided in Appendix \ref{app:BSFT}.
Instead, 
we here give another explanation with the same conclusion
in terms of projective modules
in order to avoid the unnecessary complication.
For the sake of simplicity, let us first focus
on the projection operator
$P_{N/S}(x)$ in \eqref{str of potential}
in a more general context.
The effect of the delta function is taken into account afterward.

The Chan-Paton bundle on $N$ non-BPS D3-branes
is a complex vector bundle $E$ over $\R^4$,
whose typical fiber $\C^N$ is a Hilbert space.
Let us write the orthonormal basis of $\C^N$ as
$\Ket{a}$ ($a=0,1,2,\ldots,N-1$)  such that
$\Bracket{a}{b}=\delta_{ab}$.
Let ${\cal A}=C^\infty (\R^4)$ be an algebra of smooth functions
and ${\cal A}^N$ be a free module of rank $N$,
which is the space of sections of the trivial vector bundle $\R^4\times \C^N$.
An element of this module is written as
\bea
\Ket{\psi}=\sum_{a=0}^{N-1} \psi^a (x) \Ket{a}, \quad \psi^a(x)\in {\cal A}.
\ena
We next consider a projection operator $P \in M_N({\cal A})$,
that is, $P^\dagger =P$ and $P^2=P$.
It picks up a subspace at each fiber.
This defines a projective module $P{\cal A}^N$, which is a right ${\cal A}$-module.
In general, the space of section of any vector bundle $E$ over $\R^4$
is constructed in this manner.
An element of $P{\cal A}^N$ (i.e., a local section of $E$) is written by $\Ket{\psi}$ as
\bea
\Ket{\xi}=P\Ket{\psi}=\sum_{a,b} \Ket{a}{P^a}_b(x)\Bracket{b}{\psi}
=\sum_{a,b} \Ket{a}{P^a}_b(x)\psi^b (x).
\ena
Since $P$ depends on $x$ in general, the exterior derivative $d$ does not preserve the module $P{\cal A}^N$.
This leads to the notion of connections.
A connection on $P{\cal A}^N$ is a linear map
$\nabla : P{\cal A}^N \to P{\cal A}^N \otimes_{\cal A} \Omega^1({\cal A})$
such that $\nabla (\Ket{\xi}f)=\nabla \Ket{\xi} f +\Ket{\xi}\otimes df$ for all $\Ket{\xi}$ and $f \in {\cal A}$.
A natural connection on $P{\cal A}^N$, called the Grassmannian connection, is defined by $\nabla=P\circ d$,
\bea
\nabla \Ket{\xi}=Pd \Ket{\xi}=Pd(P\Ket{\psi})
=P \Ket{d\psi} +(PdP)\Ket{\psi},
\ena
where $\Ket{d\psi}=\sum_a (d\psi^a)\Ket{a}$.
The last term $PdP$ plays the role of a gauge field%
\footnote{{
More generally, any connection on a projective module $P{\cal A}^N$
has the form $\nabla =Pd +\omega$,
where $\omega \in {\rm End}_{\cal A}(P{\cal A}^N)\otimes \Omega^1({\cal A})$,
which satisfies $P\omega P=\omega$.
The curvature in this case is $\nabla^2 \Ket{\xi}
=\left(PdPdP+ P(d\omega +\omega^2)P \right) \Ket{\psi}$.}}.
{The curvature of this connection is given by
\bea
\nabla^2 \Ket{\xi}
&=Pd \left(P\Ket{d\psi}+PdP \Ket{\psi}\right) \nn\\
&=PdP \Ket{d\psi}+PdPdP\Ket{\psi}-PdP\Ket{d\psi} \nn\\
&=PdPdP\Ket{\psi}\nn\\
&=-(dPdP) \Ket{\xi},
\ena
where we used an identity $PdP P=0$ follows from differentiating $P^2=P$.}

To be more specific, we assume that
the projection operator is rank $1$ having the form
\bea
P(x)=U(x)\Ket{0}\Bra{0}U^\dagger (x),
\ena
which is unitary equivalent to $\Ket{0}\Bra{0}$.
Here the $x$-dependence is only in the unitary operator $U(x)$.
This means that $P(x)$ picks up
a $1$-dimensional subspace $U(x)\Ket{0}$ of each fiber $\C^N$ at $x$.
Since $U$ is a unitary operator, the set $\{U(x)\Ket{a}\}$ forms an orthonormal basis.
We may then write a generic element of the free module ${\cal A}^N$ in this new basis as
\bea
\Ket{\psi}=\sum_a \psi^a (x) U(x)\Ket{a}.
\ena
An element of the projective module $P{\cal A}^N$ is then given by
\bea
P\Ket{\psi}
&=\sum_a \psi^a (x) PU(x) \Ket{a}
=\sum_a \psi^a (x) U(x) \Ket{0} \Bracket{0}{a}
=\psi^0 (x) U(x) \Ket{0}.
\ena
Since only
the $a=0$ component appears, this module $P{\cal A}^N$ is
a line bundle. This bundle is in general nontrivial
because of the $x$-dependence of $U(x)$,
which is indicated by the Grassmannian connection:
\bea
Pd(P\Ket{\psi})
&=Pd (\psi^0 U\Ket{0}) \nn\\
&=P(d\psi^0 U\Ket{0}+\psi^0 dU \Ket{0}) \nn\\
&=d\psi^0 U\Ket{0} +\psi^0 U\Ket{0}\Bra{0}U^\dagger dU \Ket{0} \nn\\
&=\left(d\psi^0 +iA \psi^0 \right) U\Ket{0},
\ena
where we have defined
\bea
iA(x)=\Bra{0}U^\dagger dU\Ket{0}.
\label{def of gauge field}
\ena
Thus, in components, we obtain
the covariant exterior derivative
$\psi^0 \to d\psi^0 +iA \psi^0$
on the line bundle with a $U(1)$ gauge potential $A$.

Next, we consider the effect of the delta function.
We denote $\delta(M)$ as a delta-function distribution
whose support is a submanifold $M \subset \R^4$.
When acting on ${\cal A}=C^\infty (\R^4)$,
it confines ${\cal A}$ to $C^\infty (M)$.
Then a projective module of the form $P {\cal A}^N \delta(M)$ becomes
a vector bundle over $M$.
In this case, the Grassmannian connection is modified schematically as
\bea
d(P\Ket{\psi}\delta(M))
&=d(P\Ket{\psi})\delta(M) +P\Ket{\psi}d_N \delta(M)  \nn\\
&=(d_M +d_N) (P\Ket{\psi}) \delta(M) -d_N (P\Ket{\psi}) \delta(M) \nn\\
&=d_M (P\Ket{\psi})\delta(M),
\label{d P delta}
\ena
where the exterior derivative $d$ has been divided into
the longitudinal and transverse components along $M$ as
$d=d_M +d_N$.
Since $d(\delta(M))=d_N(\delta(M))$,
the exterior derivative is consistently restricted to $d_M$.
It guarantees that the gauge potential in this case is just the restriction of the domain of
\eqref{def of gauge field} to $M$ as a local $1$-form on $M$.

Coming back to our present case, the trivial Chan-Paton bundle
$\R^3\times \C^2$ 
corresponds to a free module ${\cal A}^2$.
The projection operator $P_{N/S}(x)$ in \eqref{P} picks up
a $1$-dimensional subspace
generated by $U_{N/S}(x)\Ket{0}=\bv_{N/S}$ at each fiber $\C^2$
within the open set ${\cal U}_{N/S}$.
Of course, \eqref{def of gauge field} reduces the monopole potential
\eqref{monopole potential} in this case.
In addition, the delta function in \eqref{str of potential} restricts the base space
$\R^3$ to $S^2$,
and the projective module means a non-trivial Chan-Paton bundle
for a single spherical D2-brane.
As argued around \eqref{d P delta}, the 
potential \eqref{monopole potential} is obtained directly
by using the exterior derivative $d_{S^2}$.
In fact, by writing $\bv_N$ in \eqref{unitary matrix}
in the polar coordinate as
\bea
\bv_N=\frac{1}{\sqrt{2(1+\cos\theta)}}\vect{1+\cos\theta,\sin\theta e^{i\phi}},
\ena
we can explicitly check that the quantity
$\bv_N^\dagger d_{S^2} \bv_N$ with
$d_{S^2}=d\theta \partial_\theta +d\phi \partial_\phi$
coincides with $A_N$ in \eqref{monopole potential}.

\subsubsection{Remarks}

Before closing this section, let us make some remarks.

\paragraph{$U(1)$ gauge symmetry}
To diagonalize the tachyon profile, we have used an $SU(2)$ gauge transformation
$U_{N/S}$ in \eqref{unitary matrix}.
But any other eigenvectors ${\bf v}_{N/S}^\alpha =e^{i\alpha} {\bf v}_{N/S}$
and ${\bf w}_{N/S}^\alpha=e^{i\alpha} {\bf w}_{N/S}$
multiplied by an arbitrary phases $\alpha(x)$ can also be used to diagonalize $T$.
Then, the induced gauge potential is modified to
${\bf v}_\alpha^\dagger d{\bf v}_\alpha = {\bf v}^\dagger d{\bf v}+ d\alpha$
by a pure gauge factor.
Thus there is a spectator $U(1)$ degrees of freedom,
which becomes a $U(1)$ gauge symmetry on a spherical D2-brane.

\paragraph{Equation of motion}
Our profile \eqref{profile} is indeed a solution of the equation of motion
in the limit $u \to\infty$.
To see this, recall that the action \eqref{D3 action} has the structure,
\begin{align}
S_{\rm NSNS} \sim \int {\rm Tr}_2 \, V(T) {\cal F}(\partial T),
\end{align}
where $V(T)=e^{-T^2}$ is the tachyon potential and ${\cal F}(\partial T)$ is
the kinetic term, which is a specific functional of the first order derivative of $T$,
but we do not need its precise form.
Under any variation $\delta T$, the variation of the action is written,
ignoring the operator ordering, as
\begin{align}
\delta S_{\rm NSNS}
&\sim \int {\rm Tr}_2 \left[\delta T V'(T) {\cal F}(\partial T)
+ V(T)\partial (\delta T) {\cal F'}(\partial T)\right]\nn\\
&=\int {\rm Tr}_2 \delta T \left[ V'(T) {\cal F}(\partial T)
- V'(T)\partial T{\cal F'}(\partial T)- V(T)\partial^2 T{\cal F''}(\partial T)\right].
\end{align}
{The point is} that the equation of motion for $T$
contains either {$\partial^2 T$ or $V'(T)$}.
The {former} vanishes obviously for our linear tachyon profile,
while the {latter} vanishes because $V'(T)=-2T e^{-T^2}$ and
$e^{-T^2}$ picks up the kernel of $T$ in the limit $u\to\infty$.
The operator ordering does not affect this statement.

\paragraph{Topological soliton}
Let us discuss our tachyon profile in the context of topological soliton.
First we focus on the upper-left part $t(x)=u(r-R)$
of the diagonalized tachyon profile {\eqref{diagonal T}}.
We argue that this profile becomes a kink type soliton
in the radial direction in the limit $u\to \infty$.

To see this, let us briefly recall a kink in the $1+1$ dimensional real scalar field theory
${\cal L}=\frac{1}{2}\partial_\mu \phi \partial^\mu \phi-V(\phi)$
with a potential $V(\phi)=e^{-\phi^2}$.
Since it has minima at $\phi=\pm \infty$,
a kink should satisfy the boundary condition $\phi(-\infty)=-\infty$ and
$\phi(\infty)=\infty$.
A typical solution is given by $\phi(x)=ux$ ($u\to \infty$).
In this case, the topological sector is indeed unaffected by the value of the parameter $u$,
but the condition $u\to \infty$ is needed in order
to satisfy the equation of motion.
Any other profile by deforming this profile continuously with keeping the same boundary condition lies in the same topological sector.
However, when considering a scalar field in a box $x\in [-L,L]$,
the boundary condition for a kink will be changed to $\phi(-L)=-\infty$ and $\phi(L)=\infty$,
and thus
the profile $\phi(x)=ux$ satisfies this condition only in the limit $u\to \infty$.
In the latter case, another profile $\phi(x)=u\tan \left(\frac{\pi}{2L}x\right)$
would be a better one as a representative, since $u$ is irrelevant for specifying
the topological sector.

With this in mind, it is now obvious that the condition $u\to \infty$ is essential
in order that $t(x)=u(r-R)$ satisfies
the boundary condition $t(r=0)=-\infty$ and $t(r=\infty)=\infty$ for a radial kink,
where the boundary $r=0$ plays the same role as a finite box.
In other words, $u <\infty$ and $u=\infty$ can be
two distinct topological sectors.
We will come back to this point when considering
the RR-coupling of the system.

The original tachyon profile \eqref{profile} is matrix valued,
and its asymptotic behavior at $r\to\infty$
is the same as that of {the} single D0-brane solution \eqref{D0 profile},
a codimension three soliton in $\R^3$.
However, the behaviors at $r= 0$ are different:
$T=-uR 1_2$ in \eqref{profile} and $T=0$ in \eqref{D0 profile}.
They can be again in different topological sectors in the limit $u=\infty$.
At the origin, the non-BPS D3-branes annihilate in the former case,
while they stay in the false vacuum in the latter case.
This observation suggests that 
it {is} possible to distinguish the spherical D2-brane from the single D0-brane
{by their} topological sectors.

\section{Effective theory on a spherical D2-brane}

In this section, we read off the energy (tension) and the RR-coupling 
of the system
by inserting the tachyon profile \eqref{profile}
into the effective action for $2$ non-BPS D3-branes \eqref{D3 action}.
The result shows that our system is a bound state of a $D2$-brane and a $D0$-brane.
We then obtain the effective action for a spherical D2-brane
and give an analysis on the stability of the system in some RR-backgrounds.

\subsection{RR-sector}

The RR-coupling of $N$ non-BPS $D3$-branes in type IIA theory
is described by the CS-term \cite{Kraus2001}
(but here we adopt the convention used in \cite{Myers1999}),
\begin{align}
S_{\rm CS} = \sqrt{2}\mu_3 \int_{\R^4} P\left[\sum C^{(n)}\right] \wedge {\rm Tr}_N
\,\, e^{\lambda (-T^2 +DT+F)},
\label{N D3 CS}
\end{align}
where $\mu_3=\frac{2\pi}{g_s (2\pi\sqrt{\alpha'})^4}$,
$\lambda =2\pi\alpha'$,
$F=\frac{1}{2}F_{\mu\nu}dx^\mu \wedge dx^\nu$ is a $U(N)$ field strength with
$F_{\mu\nu}=\partial_\mu A_\nu -\partial_\nu A_\mu +i[A_\mu,A_\nu]$,
$DT=dT+i[A,T]$ is the covariant derivative of the tachyon field,
$C^{(n)}$ is a RR $n$-form ($n$: odd)
in the spacetime and $P[C]$ is its pullback to the worldvolume $\R^4$.
For D$3$-branes, it is sufficient to consider
the $1$-form $C^{(1)}=C_\mu dx^\mu$
and the $3$-form
$C^{(3)}=\frac{1}{3!}C_{\mu\nu\rho}dx^\mu \wedge dx^\nu \wedge dx^\rho$.
Note that only the odd numbers of $DT$ contribute to the integral.

Our case in \eqref{D3 action} is $N=2$ with the vanishing gauge field $A=0$
and the trivial pull-back $P$.
Then the regular way to evaluate the CS-term is to
insert the tachyon profile \eqref{profile} into \eqref{N D3 CS}.
{
Equivalently, 
we can insert the profile \eqref{diagonal T}
after the $SU(2)$ gauge transformation by $U_{N/S}$
given in \eqref{unitary matrix} 
since the CS-term is gauge invariant.
However
we already know
that the tachyon condensation picks up the upper-left component and that
the monopole gauge flux appears
from \eqref{str of potential} and {the} subsequent arguments.} 
Therefore, a quick way to achieve the result is to consider effectively
a single non-BPS D3-brane
and to insert a tachyon profile and the $U(1)$ monopole flux,
\begin{align}
&T=u(r-R), \quad
F=\frac{1}{2} \sin\theta d\theta \wedge d\phi,
\label{eff profiles}
\end{align}
into the action \eqref{N D3 CS} with $N=1$:
\begin{align}
S_{\rm CS}=\sqrt{2}\mu_3 \int_{\R^4} P[C^{(1)}+C^{(3)}]
\wedge e^{\lambda(-T^2+dT+F)}.
\label{eff CS}
\end{align}
Although the gauge transformation is well-defined only locally,
this expression is valid globally.
Since the theory is abelian, no trace is needed and $DT=dT$.

Recalling that the term with
$dT\wedge dT \wedge dT$ vanishes because $dT=u dr$,
the CS-term is expressed as
\begin{align}
S_{CS} &= \sqrt{2}\mu_3 \int_{\R^4} \left(
P[C^{(1)}]\wedge \lambda F \wedge \lambda dT
e^{-\lambda T^2} +P[C^{(3)}] \wedge \lambda dT e^{-\lambda T^2 }
\right).
\end{align}
Since the term including the tachyon reduces to the radial delta function,
\begin{align}
\lim_{u\to\infty}\lambda dT e^{-\lambda T^2 }
=\sqrt{\lambda \pi} dr \delta(r-R),
\end{align}
the worldvolume after the integration over the radial direction
becomes $\R\times S^2$:
\begin{align}
\lim_{u\to \infty} S_{CS} &= \sqrt{2\pi \lambda }\mu_3 \int_{\R\times S^2}
P[C^{(1)}]\wedge \lambda F +P[C^{(3)}] \nn\\
&= \mu_2 \int_{\R\times S^2}
P[C^{(1)}+C^{(3)}] \wedge e^{\lambda F},
\label{D2 CS}
\end{align}
where $\mu_2 = \sqrt{2\pi \lambda }\mu_3$ {has been} used.
Notice that $P$ means now the pull-back to $\R\times S^2$.
Therefore, the tachyon condensation produces the CS-term
for a spherical D2-brane with a unit magnetic flux $F$ on it.
Here the electric coupling to $C^{(3)}$ shows that the system has the unit
D2-brane charge with respect to the component $C^{(3)}_{0\theta\phi}$.
On the other hand, the coupling to $C^{(1)}$ shows the system carries
the unit D0-brane charge.
Indeed, if $C^{(1)}$ is constant along $S^2$, we have
\begin{align}
\mu_2 \int_{\R\times S^2}
P[C^{(1)}] \wedge \lambda F
&=\mu_0 \int_{\R}  C^{(1)}_0 dt, 
\label{RR1coupling}
\end{align}
where $\mu_0=2\pi \lambda \mu_2$.

\subsubsection{Direct computation}
\label{subsec:direct}
It is {of course} possible to evaluate \eqref {N D3 CS} {for $N=2$} {honestly},
{by inserting} the profile \eqref{diagonal T} after the $SU(2)$ gauge transformation $U_{N/S}$ \eqref{unitary matrix}%
\footnote{
We write formulae in the open set ${\cal U}_N$ without subscript $N$,
but the computations in ${\cal U}_S$ is completely parallel.}.
The price to pay is that
the pure gauge potential $iA=U^\dagger d U$
enters in the covariant derivative $DT$ in \eqref {N D3 CS}, which can be written as
\begin{align}
DT= u \left(\begin{matrix}
dr & -2r\bv^\dagger d \bw \\
2r \bw^\dagger  d \bv & -d r
\end{matrix}\right).
\end{align}
Note that field strength $F$ is still absent in \eqref {N D3 CS}
due to the pure gauge connection.
By using the formula,
\begin{equation}
e^{A+B}=e^A\, {\rm P} e^{\int_0^1 ds B(s)}, \quad
B(s) \equiv e^{-sA} B e^{sA},
\label{formula}
\end{equation}
we can expand the exponential in \eqref {N D3 CS} in powers of $DT$, 
where ${\rm P}$ stands for the path ordered symbol.
Then, \eqref {N D3 CS} can be written as
\begin{align}
S_{\rm CS} =
& \sqrt{2} \mu_{3} \int_{\R^4} \biggl\{
\lambda C^{(3)} \wedge
{\rm Tr}_2 \left( \int_0^1 \!\!ds\, DT(s)\, e^{-\lambda T^2} \right) \nn \\
&+\lambda^3 C^{(1)} \wedge
{\rm Tr}_2 \biggl(  \int_0^1 \!\!ds_1  \int_0^{s_1}\!\!ds_2  \int_0^{s_2} \!\!ds_3\,
DT(s_1) \wedge DT(s_2) \wedge DT(s_3)\,
 e^{-\lambda T^2}
\biggr)\biggr\},
\label{CS action2}
\end{align}
with
\begin{align}
&DT(s)= u \left(\begin{matrix}
dr & -2r e^{ -4\lambda u^2 R r s }\, \bv^\dagger d \bw \\
2r e^{4\lambda u^2 R r s}\, \bw^\dagger  d \bv & -d r
\end{matrix}\right).
\end{align}
We can easily see that this expression gives the same results 
\eqref{D2 CS} {and \eqref{RR1coupling}}
in the limit $u\to \infty$.
In particular, the monopole flux comes from the cubic term of $DT(s)$ 
in \eqref{CS action2}.

\subsubsection{D-brane density in the spacetime}
In the above computations of the RR-coupling, we assume the limit $u\to \infty$.
However,  in the familiar examples of the tachyon condensations,
the CS-term is independent of the condensation parameter $u$.
This is accordance with the well-known result
that the D-brane charge is topological and must
be an integer \cite{Witten1998}.
Here we examine the $u$-dependence of our result \eqref{CS action2} more closely.

Recall that a D-brane system can in general couple to multiple RR
potentials and the couplings are expressed in the form of the 
space-time integral as
\begin{equation}
S_{\rm RR} = \int_{\R^{10}} \sum_{p} C^{(p+1)}(x) \wedge \rho^{(9-p)}(x),
\end{equation}
where $p$ runs even(odd) numbers for the Type IIA(IIB) superstring theory and
$C^{(p+1)}$ is the RR $(p+1)$-form potential.
The $(9-p)$-form $\rho^{(9-p)}$ expresses
the ``D$p$-brane density'' of the system,
which acts as a source in the equation of motion
of the RR $(p+1)$-form potential,
\begin{equation}
d*dC^{(p+1)}(x) = \rho^{(9-p)}(x).
\end{equation}
For example, a BPS D$p$-brane sitting at the origin
in its transverse $\R^{9-p}$ corresponds to
$\rho^{(9-p)}=\mu_p \delta^{p+1}(x^{p+1},\cdots,x^9)
dx^{p+1}\wedge \cdots \wedge dx^9$.
We do not solve this equation of motion explicitly since we
need to solve simultaneously with the Einstein equation {in general}.
We instead focus on the D$p$-brane density of our system.

The D2-brane density coupling to $C^{(3)}$ can be read off 
from the first line
of (\ref{CS action2}) as
\begin{equation}
\rho^{(7)}=\sqrt{2}\mu_3 \lambda u
\left( e^{-\lambda u^2 ( r-R )^2 } - e^{-\lambda u^2 ( r+R )^2} \right)
\delta^{(6)}(x^4,\cdots,x^9) dr \wedge dx^4 \wedge\cdots \wedge dx^9, 
\label{D2 density}
\end{equation}
which is distributed along a spherical thick shell in $\R^3$
which has a peak at $r=R$ and vanishes at the origin $r=0$.
It reduces to the delta-function distribution in the limit $u\to\infty$,
\begin{equation}
\lim_{u\to\infty} \rho^{(7)} =
\mu_2 \, \delta(r-R)
\, \delta^{(6)}(x^4,\cdots,x^9) \,
dr \wedge dx^4 \wedge\cdots \wedge dx^9,
\end{equation}
which is localized at $r=R$ and $x^4=\cdots=x^9=0$.
Thus, by integrating the latter over $\R^+\times \R^6$ given by
fixing $t$, $\theta$ and $\phi$, we obtain the unit D2-brane charge $\mu_2$ in this limit.
This is nothing but the Gauss law for RR $3$-form potential $C^{(3)}$.
For a finite $u$, on the other hand, the same integral gives
\begin{align}
\int_{\R^+\times \R^6}\rho^{(7)}
&=\sqrt{2}\mu_3 \lambda u
\int_0^\infty
\left( e^{-\lambda u^2 ( r-R )^2 } - e^{-\lambda u^2 ( r+R )^2} \right)
dr \nonumber \\
&= \mu_2\,  {\rm erf}(\sqrt{\lambda}uR ),
\label{D2 charge1}
\end{align}
where
${\rm erf}(x)=\frac{2}{\sqrt{\pi}} \int_0^x dt e^{-t^2}$ is the error function.
It depends continuously on $uR$, and
reduces to $\mu_2$ only in the limit of $u\to\infty$. 
It shows that the tachyon condensation in our case is not a familiar type
mentioned above.
However, it is in accordance with the charge quantization
since our tachyon profile can be regarded as 
a topological soliton in the region $0\le r <\infty$
 only in the limit of $u\to\infty$
as argued in the previous section.

Similarly, the D0-brane density coupling to $C^{(1)}$, 
which can be read off from the second line
of (\ref{CS action2}), is
\begin{equation}
\rho^{(9)}=
\frac{\sqrt{2}\mu_3 \lambda^2 u}{2R}
\frac{ e^{-\lambda u^2( r-R )^2 } - e^{-\lambda u^2( r+R )^2}}{r}
\delta^{(6)}(x^4,\cdots,x^9)\,
dx^1\wedge \cdots \wedge dx^9.
\label{D0 density}
\end{equation}
This distribution is slightly different from that of D2-brane density \eqref{D2 density}.
In particular, $\rho^{(9)}\ne 0$ at $r=0$.
However, \eqref{D0 density} also approaches to the radial {delta-function} distribution
in the limit $u\to \infty$.
This shows that a single D0-brane is resolved into $S^2$.
It is compared to the D0-brane density $\rho^{(9)}_{\rm D0}$
coming from the D0-brane solution \eqref{D0 profile},
\begin{align}
\rho^{(9)}_{\rm D0}
&=\sqrt{2}\mu_3 2(\lambda u)^3
e^{-\lambda u^2|x|^2 }
\delta^{(6)}(x^4,\cdots,x^9)\,
dx^1\wedge \cdots \wedge dx^9 \nn\\
&\xrightarrow{u\to\infty}
\mu_0 \delta^{(9)}(x^1,\cdots,x^9)\,
dx^1\wedge \cdots \wedge dx^9,
\label{D0' density}
\end{align}
which approaches to the {delta-function} distribution at the origin.
Integrating $\rho^{(9)}$ over the space $\R^9$, we have
\begin{align}
\int_{\R^9}\rho^{(9)}
&= \frac{\sqrt{2}\mu_3 \lambda^2 u}{2R}
\int_{\R^{3}}
\frac{ e^{-\lambda u^2( r-R )^2 } - e^{-\lambda u^2( r+R )^2}}{r}
dx^1 dx^2 dx^{3} \nonumber \\
&= \mu_0,
\end{align}
which is the same as the integration of $\rho^{(9)}_{\rm D0}$.
Therefore, this system has a unit D0-brane charge
irrespective of the values of $u$ and $R$.

\subsection{NSNS-sector}
{In order to identify the energy of the system, we} 
insert the tachyon profile \eqref{profile}
into the NSNS-part of the effective action for $2$ non-BPS D3-branes.
Apart from the BSFT action \eqref{D3 action},
there are many {different forms of the action 
related} with each other by field redefinitions. 
We {here adopt} a Dirac-Born-Infeld (DBI) like action among them.
By the same reasoning in the RR-sector, it is sufficient effectively to
insert \eqref{eff profiles} into the action for a single non-BPS D3-brane
(in the static gauge),
\begin{align}
S_{\rm DBI} =-\sqrt{2}T_3 \int_{\R^4} d^4 x
e^{-\lambda T^2} \sqrt{ \det_{\mu,\nu} \left(
-\eta_{\mu\nu} + \lambda F_{\mu\nu} + \lambda^2 \del_\mu T\,\del_\nu T
\right)}.
\label{D3 DBI}
\end{align}
In the polar coordinates $x^\mu=(t,r,\theta,\phi)$,
$\eta_{\mu\nu}$ corresponds to the line element
$ds^2=-dt^2 +dr^2+r^2(d\theta^2 +\sin^2\theta d\phi^2)$.
For the profile \eqref{eff profiles}, the non-vanishing components are
$F_{\theta\phi}$ and $\partial_r T \partial_r T= \lambda^2 u^2$, respectively.
Thus, the determinant factor can be written as
\begin{align}
\det_{\mu,\nu} \left(
-\eta_{\mu\nu} + \lambda F_{\mu\nu} + \lambda^2 \del_\mu T\,\del_\nu T\right)
&=(1+\lambda^2 u^2) \det_{a,b} (g_{ab}+\lambda F_{ab}),
\end{align}
where $x^a=(\theta,\phi)$ is the angular part and
$g_{ab}$ corresponding to $ds^2= r^2(d\theta^2 +\sin^2\theta d\phi^2)$
is the metric on $S^2$ with radius $r$.
By using this, the DBI-like action reduces in the $u\to\infty$ limit to
\begin{align}
\lim_{u\to \infty} S_{\rm DBI}
&=-\lim_{u\to \infty}\sqrt{2}T_3 \int_{\R^4} dt dr d\theta d\phi \,
e^{-\lambda u^2 (r-R)^2}
\sqrt{1+\lambda^2 u^2} \sqrt{\det_{a,b} (g_{ab}+\lambda F_{ab})}\nn\\
&=-\lim_{u\to \infty}\left(\sqrt{2}T_3 \frac{1}{u}\sqrt{\frac{\pi}{\lambda}}
\sqrt{1+\lambda^2 u^2}\right)
\int_{\R\times S^2} dt d\theta d\phi
\sqrt{\det_{a,b} (g_{ab}+\lambda F_{ab})},
\label{D2 pre DBI}
\end{align}
where we have used 
$\lim_{u\to \infty}e^{-\lambda u^2 (r-R)^2}=\frac{1}{u}
\sqrt{\frac{\pi}{\lambda}}\delta (r-R)$ in integrating over $r$.
Thus, $S^2$ above is now a sphere with the radius $R$.
The factor in front of the integral reduces to the D2-brane tension:
\begin{align}
\lim_{u\to \infty}
\sqrt{2}T_3 \frac{1}{u}\sqrt{\frac{\pi}{\lambda}}\sqrt{1+\lambda^2 u^2}
&=\lim_{u\to \infty}T_2 \sqrt{1+\frac{1}{\lambda^2 u^2}}
=T_2,
\end{align}
where the relation $T_2=\sqrt{2\pi \lambda }T_3$ has been used.
Therefore, by recovering the time component in the determinant,
\eqref{D2 pre DBI} can be written as
\begin{align}
\lim_{u\to \infty}S_{\rm DBI}=-T_2
\int_{\R\times S^2} d^3 \xi \sqrt{\det_{\alpha,\beta}
(-g_{\alpha\beta}+\lambda F_{\alpha\beta})},
\label{D2 DBI}
\end{align}
where $\xi^\alpha=(t,\theta,\phi)$ and the metric $g_{\alpha\beta}$
on $\R\times S^2$ corresponds to
$ds^2=-dt^2 +R^2(d\theta^2 +\sin^2\theta d\phi^2)$.
This is nothing but the DBI action.

Note that $F_{\theta\phi}$ is arbitrary in the above computation.
For the monopole flux $F_{\theta\phi}=\frac{1}{2}\sin\theta$,
we can evaluate the rest energy of this system $E$ by
integrating over $(\theta,\phi)$ as
\begin{align}
&\lim_{u\to \infty} S_{\rm DBI} = -E \int_{\R} dt ,\quad
E=4\pi T_2 \sqrt{R^4 +\frac{\lambda^2}{4}},
\label{bound state tension}
\end{align}
which
has a typical form of the tension of a BPS bound state.
In fact, when the flux is absent ($\lambda=0$ in \eqref{bound state tension}),
it reduces to the rest energy $E \to T_2 {\rm vol}(S^2)$ for a D2-brane
wrapping a sphere with radius $R$,
while if the D2-brane is absent
(set $R=0$ formally in \eqref{bound state tension}),
it reduces to the D0-brane tension $E \to 2\pi \lambda T_2=T_0$.
The inequality $E < T_2 {\rm vol}(S^2)+T_0$ shows that the system is
more stable than the individual system of a D2-brane and a D0-brane.

\subsection{Effective action for a spherical $D2$-brane}
The results above indicate that the low energy effective action of our system is
given by the DBI-action and the CS-term action on $S^2$:
\begin{align}
&S=S_{\rm DBI}+S_{\rm CS}, \nn\\
	&S_{\rm DBI}=\bc{-}T_2 \int_{\R\times S^2} d^3 \xi \sqrt{-\det (g+\lambda F)}, \nn\\
&S_{\rm CS}=\mu_2 \int_{\R\times S^2} P[C^{(1)}+C^{(3)}]\wedge e^{\lambda F}.
\label{D2 action}
\end{align}
Here the dynamical variables of this theory
are a $U(1)$ gauge field and scalar fields which are transverse to $S^2$.
It means that we just regard $F$ in \eqref{D2 action} as dynamical 
degrees of freedom%
\footnote{More precisely, the sum of the monopole flux and a fluctuation.}.
This is possible because the derivation of this form \eqref{D2 action}
do not use the explicit form of the monopole potential.
On the other hand, the scalar fields can be incorporated in the induced
metric $g_{\alpha \beta }$ of the target spacetime%
\footnote{In principle, these fields should be added to the original
tachyon profile \eqref{profile} as fluctuations,
but the result is the same.}.  

Here we should note 
that the action \eqref{D2 action} has the same form as 
that of the dual D2-brane description of the fuzzy sphere solution 
of multiple D0-branes (the dielectric D0-branes) 
constructed in \cite{Myers1999}.
There are however two important differences 
between our system and the dielectric D0-branes.
One difference is the number of the monopoles.
Since a fuzzy sphere corresponds to an $su(2)$ representation
and the size of the representation equal to the monopole charge,
the construction in \cite{Myers1999} requires a $k$-monopole
flux ($k\ge 2$) in \eqref{D2 action}
while our monopole has a unit magnetic charge $k=1$. 
Another difference is the existence of the dual D0-brane description.
The original system of the dielectric D0-branes is made of $k$ D0-branes
and the D2-brane by the action \eqref{D2 action} is the dual picture 
corresponding to the case where the scalar fields on them have a fuzzy 
sphere configuration. 
Since the equivalence between these pictures is based on 
the matrix regularization of D2-branes, 
\eqref{D2 action} is supposed to be valid only in the large $k$ limit.
This is in contract to the fact that our tachyon profile \eqref{profile}
directly describes the bound state of a spherical D2-brane and a D0-brane.

\subsection{{Analysis on stability}}
Let us now consider the stability of the system with respect to 
the radius of $S^2$. 
Despite of the differences mentioned in the previous subsection 
we can examine the stability of the system in the same way 
argued in \cite{Myers1999}%
\footnote{See also \cite{Hashimoto2002} and \cite{Hyakutake2005}
for a similar stability analysis.}.

{In} the low energy effective action \eqref{D2 action}, 
{we consider the scalar field in the radial direction as a dynamical variable
and assume that it is constant $R(\xi)=R$.}
Plugging the metric of $S^2$ given below \eqref{D2 DBI} 
and the monopole field strength \eqref{eff profiles} into 
$S_{\rm DBI}$ in \eqref{D2 action}, we obtain 
\begin{align}
S_{DBI}{(R)}=-4\pi T_2 \sqrt{R^4 + \frac{\lambda^2 }{4}} \int_{{\R}} dt, 
\label{pot DBI}
\end{align}
which is of course identical to  \eqref{bound state tension},
{but is now considered to be a function of $R$.}

The CS-term depends on the value of the background RR fields. 
Since the coupling to the RR 1-form potential \eqref{RR1coupling}
is independent of $R$, it is nothing to do with the stability 
of the system with respect to $R$. 
We then set $C^{(1)}=0$ and consider only a background of the RR 3-form potential.

{We} first consider a constant RR 4-form field strength background, 
\begin{align}
&\left\{
\begin{array}{cl}
F^{(4)}_{0ijk}
=-2f \epsilon^{ijk} & i,j,k \in \{ 1,2,3\} \\
0 & \text{otherwise}
\end{array}
\right.. 
\label{const RR4}
\end{align}
Since the corresponding RR 3-form potential can be written as $C^{(3)}_{t\theta\phi}=\frac{2}{3}f r^3 \sin\theta$ 
in the polar coordinate, 
we can evaluate the CS-term as 
\begin{align}
	S_{\rm CS}{(R)}=\mu_2 \int_{{\R\times} S^2 } P[C^{(3)}]
&=\mu_2 \frac{8\pi}{3}fR^3 \int_{\R} dt, 
	\label{pot CS1}
\end{align}
in the background of \eqref{const RR4}.

\begin{figure}[htbp]
\begin{center}
	\includegraphics[clip,width=60mm]{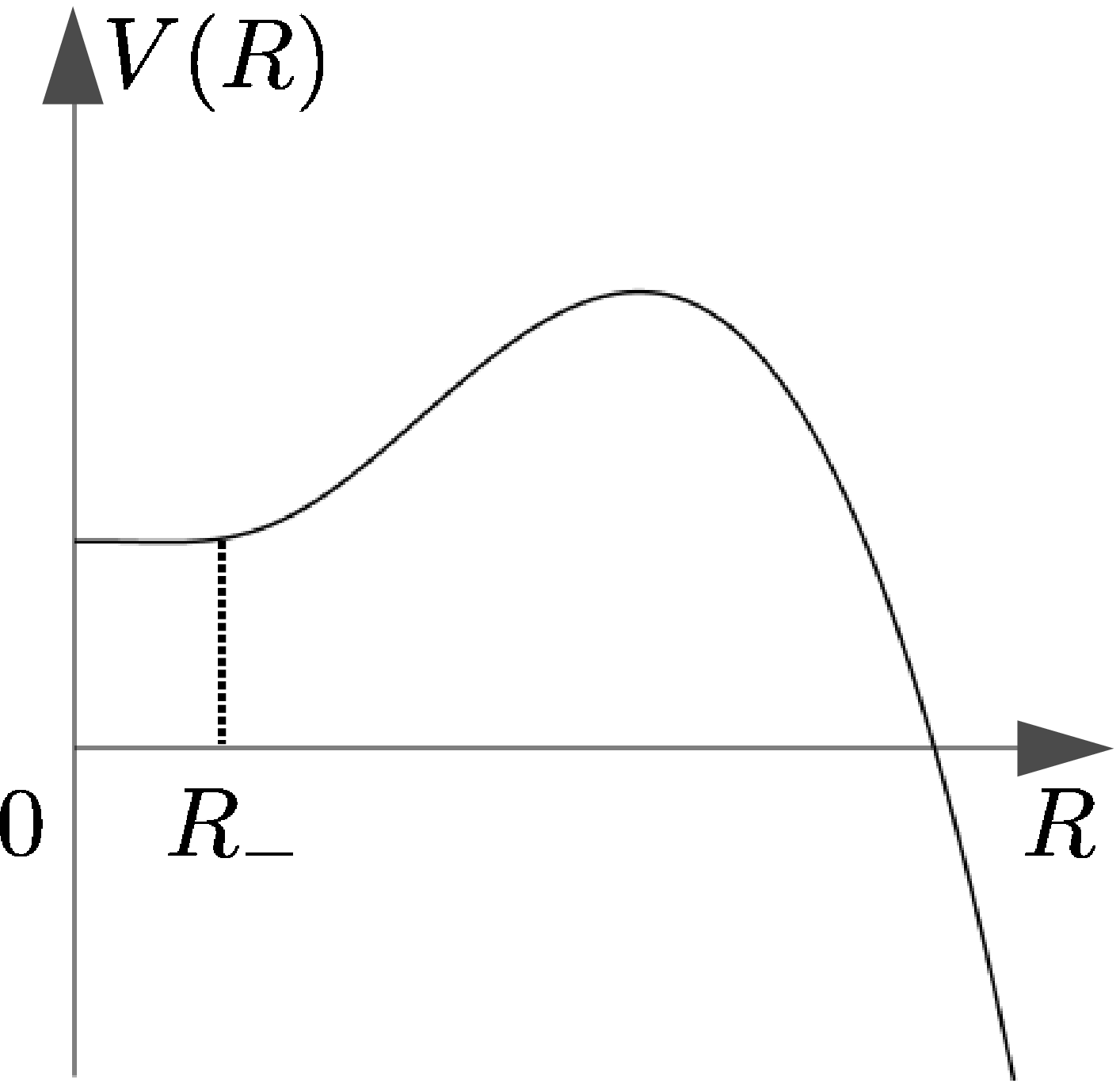}
	\includegraphics[clip,width=60mm]{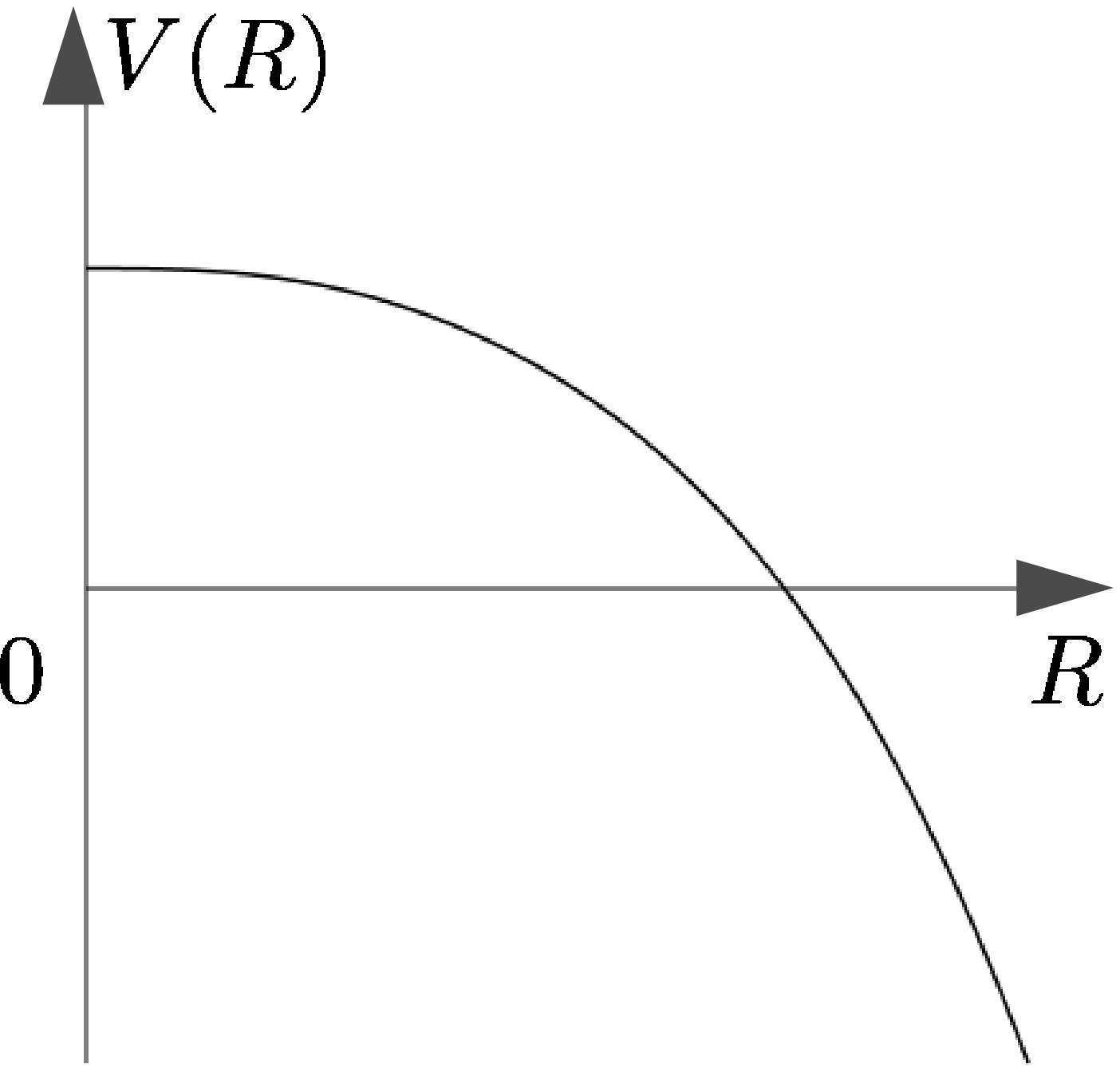}
	\caption{The shape of the effective potential \eqref{eff pot1}
		in the constant RR 4-form field strength background \eqref{const RR4}
	for $\lambda f^2 <1$ (left) and $\lambda f^2 \ge 1$ (right). 
$R_-$ in the left panel is the local minimum of $V(R)$ 
and is given by \eqref{R-}.}
\label{fig:potential1}
\end{center}
\end{figure}
Combining \eqref{pot DBI} and \eqref{pot CS1}, 
we obtain the effective potential for the transverse scalar
field in the radial direction $R$ in the constant RR 4-form field strength:
\begin{align}
V(R)&= 4\pi \sqrt{R^4 + \frac{\lambda^2 }{4}} T_2 -\mu_2 \frac{8\pi}{3}fR^3
=4\pi T_2 \left(\sqrt{R^4 + \frac{\lambda^2 }{4}} - \frac{2}{3}fR^3\right), 
\label{eff pot1}
\end{align}
whose shape is drawn in the Fig.\ref{fig:potential1}. 
When $f$ satisfies 
\begin{equation}
\lambda f^2 < 1, 
\label{stability}
\end{equation}
$V(R)$ has a local minimum at 
\begin{equation}
	R=R_-\equiv \frac{\sqrt{ 1 - \sqrt{1-\lambda^2 f^4} }}{\sqrt{2} f}, 
	\label{R-}
\end{equation}
while 
there is no local minimum 
{for $\lambda f^2 \ge 1$.}
{In particular, if $\lambda f^2 \ll 1$, 
the local minimum $R_-$ and its energy is given} approximately as
\begin{align}
{R_- \sim \frac{\lambda}{2}f, \quad}
V(R_-) \sim T_0 \left( 1- \frac{\lambda^2 f^4}{24} \right).
\end{align}
This means that the spherical D2-brane is stabilized 
at a very small radius with the size of the string length.
It is more stable than a single D0-brane ($R=0$) but 
the difference of the energy is small. 

One way to stabilize the spherical D2-brane in a macroscopic radius is 
to construct a bound state of a D2-brane and a large number $k$ 
of D0-branes as discussed in \cite{Myers1999}. 
In fact, 
when the $U(1)$ gauge flux is $F=\frac{k}{2}\sin\theta$, 
the {effective} potential for $R$ becomes 
\begin{align}
V_k (R)=4\pi T_2 \left(\sqrt{R^4 + \frac{\lambda^2 k^2}{4}} - \frac{2}{3}fR^3\right).
\label{potentialN}
\end{align}
{In \cite{Myers1999} it is assumed that}
\begin{equation}
\frac{R^2}{\lambda k} \ll 1, 
\label{condition2}
\end{equation}
{and the Taylor expansion of the effective potential is performed .  
Then,} there is a local minimum at $R_-=\frac{\lambda}{2}fk$ in the potential \eqref{potentialN}. 
{It implies $\lambda f^2 k \ll 1$}, in order that $R_-$ is in the range of \eqref{condition2}.
If we {further} impose $R_-$ to be much larger than the string scale,
{$1\ll \frac{R_-}{\sqrt{\lambda}}=\sqrt{\lambda f^2}\frac{k}{2}$, 
$k$ must be in the range}
\begin{align}
\lambda f^2 \ll \frac{1}{k} \ll \sqrt{\lambda f^2}, 
\end{align}
which is possible for sufficiently large $k$ under the condition $\lambda f^2 \ll 1$.  
Thus a macroscopic spherical $D2$-brane in the background of a constant RR 4-form 
field strength is possible {but in a delicate way. 
Our case $k=1$ is of course out of this range.}

\begin{figure}[htbp]
\begin{center}
	\includegraphics[clip,width=60mm]{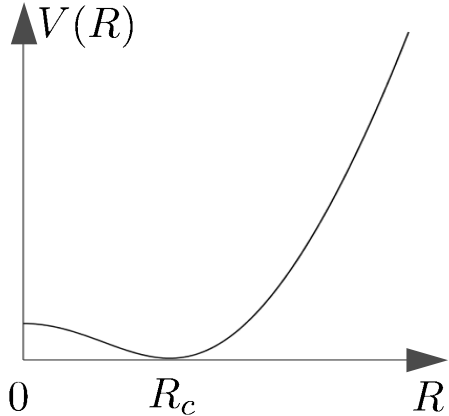}
	\caption{The shape of the effective potential \eqref{eff pot2}
		in the background of the constant RR 3-form potential \eqref{const RR3}
		with $0<C<1$. 
$R_c$ is the local minimum of $V(R)$ and is given by \eqref{Rc}.}
\label{fig:potential2}
\end{center}
\end{figure}
Another way to stabilize a macroscopic spherical D2-brane is 
to consider a background of the RR 3-form potential 
which gives weaker force than the constant flux \eqref{const RR4} 
in the large $r$ region. 
One example is 
\begin{align}
C^{(3)}_{0\theta\phi}=C r^2 \sin\theta,
\label{const RR3}
\end{align}
where $C > 0$ is a positive constant parameter.
As in the previous discussion, we obtain the {effective} potential for $R$ as
\begin{align}
V(R)&=4\pi T_2 \left(\sqrt{R^4 + \frac{\lambda^2 }{4}} - C R^2\right). 
\label{eff pot2}
\end{align}
Here we assume $C<1$ then the shape of the potential can be drawn as in Fig.\ref{fig:potential2}. 
There is a global minimum at $R_c$ given by 
\begin{align}
R_c^2 = \frac{\lambda }{2}\sqrt{\frac{C^2}{1-C^2}}. 
\label{Rc}
\end{align}
{In particular,}
if $C$ satisfies 
\begin{equation}
	C^2=1-\epsilon^2 \quad (\epsilon \ll 1),
\end{equation}
{then the radius $R_c$ and} the value of the potential becomes 
\begin{align}
{R_c \sim \frac{\lambda}{2\epsilon},\quad}
V(R_c) \sim 4\pi T_2 \frac{\lambda }{2} \epsilon = T_0 \epsilon. 
\end{align}
{Therefore,} the spherical D2-brane is stable at the macroscopic radius
{much larger than the string scale,} even if {$k=1$,
and the energy of the system is much smaller than that of the single D0-brane.}

\section{{Conclusion and discussion}}
\label{sec:conclusion}

In this paper, 
we constructed a bound state of a spherical D2-brane 
and a D0-brane from non-BPS D3-branes 
via tachyon condensation. 
The tachyon profile $T=u(\sigma\cdot x - R)$
is a simple deformation of the single BPS D0-brane
solution $T=u \sigma\cdot x$, 
which realizes the spherical worldvolume of D2-brane with the radius $R$ 
and a D0-brane is resolved into $S^2$ as a $U(1)$ monopole flux.
We checked that this system has the correct tension and the RR-coupling. 
We argued that the size of $S^2$ can be macroscopic 
in an appropriate RR 3-form potential background.

There are several issues that we did not touch upon in this paper.
First, we should understand the dynamical properties of this system more closely. 
Although we examined the stability of our system, 
the spherical D2-brane is treated as a probe in a fixed background
and we did not consider the backreaction to the bulk supergravity. 
The first step would be to find a classical solution of 
the bulk supergravity corresponding to the spherical D2-brane
with fluctuations if possible. 
Some care is needed in treating fluctuations.
For example, 
if the size of the spherical D2-brane is of the string scale, 
extra tachyonic modes coming from open strings stretching
between mutually antipodal points on $S^2$ would be admitted, 
as same as a system of D-brane and anti-D-brane 
where a tachyonic mode arises when the distance of the two branes 
is in the string scale.
If it is the case, a small spherical D2-brane would annihilate into the vacuum.
This would be related to the classification of D-branes in terms of
K-homology (see \cite{Asakawa2002} and references therein),
where the vector bundle modification seems to correspond to this process.
To study such open strings, it is useful to construct the boundary state
of this system \cite{Asakawa2003a}. 

As argued in the main text, this system is
complementary to the dielectric D2-brane considered in \cite{Myers1999}
and there are two differences in the monopole flux and the existence of the
dual D0-brane picture.
It is thus interesting to study the dielectric D-branes through our method.
One can consider higher monopole flux $k\ge 2$ in our setting.
A quick way is starting with the non-BPS D3-branes with 
$(k-1)$ monopole flux. 
The remaining $1$ monopole flux arises from the rank $1$ projection 
as described in this paper. 
Of course, 
a more systematic construction with the rank $k$ projection {would} also be possible.
{To move to the dual D0-brane picture, it is further needed to consider the matrix regularization of $S^2$,}
which will be studied in a future publication. 

In this paper, we concentrated on the construction of a spherical D2-brane,
but it is also interesting to generalize it to construct other
D-brane bound states with curved worldvolumes through tachyon condensation.
A generalization to higher dimensional spheres is rather straightforward: 
We can consider the same deformation of the ABS construction,
which will give a bound state of spherical D-brane(s) and a D0-brane. 
In any case, information of the resulting system is encoded 
in the tachyon profile. 
The description given in this paper in terms of
projective modules (with delta distribution) is quite general and
is applicable to other situations.
Here the projection carries the information on a non-trivial
gauge connection
and the delta distribution represents a curved worldvolume. 
We hope to report other examples along this description in the near future.

\section*{Acknowledgments}
The authors would like to thank
Poul Henrik Damgaard,
Shunji Matsuura {and Satoshi Watamura}
for useful discussion.
S.~M. would like to thank people in Niels Bohr Institute
for their hospitality in his stay.
The work of S.~M. is supported in part by Grant-in-Aid for Scientific Research (C) 15K05060.

\appendix
\setcounter{equation}{0}
\section{Radial $\delta$-function}
\label{app:radial delta}
In the following, we show that the function,
\begin{equation}
{\delta(r-R)}\equiv \lim_{u\to\infty} \frac{u^{2m+1}}{\Gamma(m+1/2)}
(r-R)^{2m} e^{-u^2 (r-R)^2 }
\label{gen delta}
\end{equation}
satisfies
\begin{equation}
\int_0^\infty dr {\delta(r-R)} f(r) = f(R),
\label{delta def}
\end{equation}
for an arbitrary function $f(r)$ and a non-negative integer $m$.

To show it, let us consider the integration,
\begin{align}
I(R,u)&\equiv \int_0^\infty dr
\frac{u^{2m+1}}{\Gamma(m+1/2)}
(r-R)^{2m} e^{-u^2 (r-R)^2 } f(r).
\end{align}
By defining
\begin{equation}
t \equiv {u(r-R)},
\end{equation}
we can evaluate the integration $I(R,u)$ as
\begin{align}
I(R,u)&=\frac{1}{\Gamma(m+1/2)}
\int_{-uR}^{\infty} dt \,
t^{2m} e^{-t^2} f(R+t/u) \nonumber \\
&=\frac{1}{\Gamma(m+1/2)}
\int_{-uR}^{\infty} dt \,
\sum_{n=0}^\infty \frac{t^{2m+n}}{u^n} e^{-t^2} f^{(n)}(R) \nn \\
&= \frac{1}{\Gamma(m+1/2)} \sum_{n=0}^\infty
\frac{f^{(n)}(R)}{u^n}  \nn \\
&=\frac{1}{\Gamma(m+1/2)} \sum_{n=0}^\infty \frac{ f^{(n)}(R) }{2 u^n}
\left( \Gamma(m+(n+1)/2) + \gamma(m+(n+1)/2, u^2R^2) \right),
\end{align}
where $\gamma(s,x)\equiv \int_0^x t^{s-1} e^{-t}$
is the lower incomplete gamma function.
From this expression, it is easy to see that only the contribution
from $n=0$ survives and
\begin{equation}
\lim_{u\to\infty} I(R,u) =  f(R)
\end{equation}
which is nothing but the relation (\ref{delta def}).

\section{BSFT analysis}
\label{app:BSFT}

\subsection{Review of BSFT description of tachyon condensation}
\label{app:review}

Let us start with a brief review of the tachyon condensation
via the boundary string field theory (BSFT) in Type II superstring theory
\cite{Witten1992c,Witten1993a,Shatashvili1993a,Shatashvili1993}.
We here consider only the excitation of the tachyon field
$T(x)$ and the gauge fields $A_\mu(x)$, which are both $N\times N$ hermitian matrices,
and do not consider the massless scalar fields and the massive modes
for simplicity.

The BSFT action and the CS-term
for this system
are given as a disk partition function with a boundary interaction $S_b[T,A_\mu]$
\cite{Kraus2001,Takayanagi2000}
acting on the boundary state of the NSNS sector%
\footnote{
The normalization factor is determined from
$\frac{2\pi \sqrt{2}}{g_s} \int [d\bX^\mu] \langle 0 | \bX^\mu \rangle_{\rm NS}
= \sqrt{2}T_{p} \int d^{p+1}x$,
where $T_{p}=[(2\pi)^p \alpha^{\prime \frac{p+1}{2}} g_s]^{-1}$
is the tension of a single BPS D$p$-brane.
The factor $\sqrt{2}$ is necessary to reproduce the tension of
a non-BPS D$p$-brane.
See \cite{Asakawa2002} for detail.
},
\begin{equation}
S(T,A_\mu ) =  \frac{2\pi \sqrt{2}}{g_s} \langle 0 | e^{-S_b[T,A_\mu]} | Bp \rangle_{\rm NS},
\label{BSFT action}
\end{equation}
and the RR sector,
\begin{equation}
S_{\rm CS}(C,T,A_\mu ) =  \sqrt{2} \mu_p \langle C | e^{-S_b[T,A_\mu]} | Bp \rangle_{\rm RR},
\label{CS term}
\end{equation}
respectively,
where
$g_s$ is the string coupling constant,
$| 0 \rangle$ is the vacuum state of the NSNS sector,
$\mu_p$ is the RR $(p+1)$-form charge of a single BPS D$p$-brane,
$| C \rangle$ is the state coupling to a RR field
\cite{Kutasov2000,Kraus2001,Takayanagi2000},
and
$|Bp\rangle_{\rm NS(RR)}$ are the boundary states given by
\begin{equation}
| Bp \rangle \equiv_{\rm NS(RR)} \int [d \bX^\mu] | \bX^\mu, \bX^a=0 \rangle_{\rm NS(RR)}.
\quad (\mu=0,\cdots,p,\  \ a=p+1,\cdots,9)
\end{equation}
Note that
$ | \bX^M \rangle_{\rm NS(RR)}$ $(M=(\mu,a))$ are the eigenstates
of the closed string superfields restricted on the boundary,
\begin{equation}
\bX^M(\hat\sigma) = X^M(\sigma) + i \theta \Psi^M(\sigma),  \quad (0\le \sigma < 2\pi)
\end{equation}
in the NSNS(RR) sector,
and $\hat\sigma$ is the boundary super-coordinate $\hat{\sigma}=(\sigma,\theta)$.
Here we have set  $\alpha'=1$ and omitted the ghost contribution and the sign $\pm$ for the spin structure, which is irrelevant in the following discussion.

The boundary interaction is given by
\begin{equation}
e^{-S_b}=\begin{cases}
\frac{1}{\sqrt{2}} {\rm Tr} \hat{P} e^{\int d\hat\sigma \bM(\hat\sigma)},  & \text{ (NSNS sector)} \\
{\rm Str} \hat{P} e^{\int d\hat\sigma \bM(\hat\sigma)},  & \text{(RR sector)} \\
\end{cases}
\label{bdr int super}
\end{equation}
with
\begin{equation}
\bM=
\left(\begin{matrix}
-i A_\mu(\bX) D\bX^\mu & T(\bX) \\
T(\bX) & -i {A}_\mu(\bX) D\bX^\mu
\end{matrix}\right),
\label{superM}
\end{equation}
where ${\rm Str}$ denotes the supertrace,
$D\equiv \del_\theta + \theta \del_\sigma$ is the
covariant derivative with respect to the supercoordinate $\hat\sigma$,
and
$\hat{P}$ denotes the supersymmetric path-ordered product defined by
\begin{equation}
\hat{P}e^{\int d\hsigma \bM(\hsigma)} \equiv
1 + \sum_{n=1}^\infty d\hsigma_1\cdots d\hsigma_n
\bM(\hsigma_n)\Theta(\hsigma_n-\hsigma_{n-1}) \bM(\hsigma_{n-1})
\cdots \Theta(\hsigma_2-\hsigma_1)\bM(\hsigma_1).
\label{super path order}
\end{equation}

It is sometimes convenient to expand $\bM(\hat\sigma)$  by $\theta$ as
\begin{equation}
\bM(\hat\sigma)=M_0(\sigma)+\theta M_1(\sigma).
\end{equation}
Then the supersymmetric path ordered product in (\ref{bdr int super})
is rewritten with respect to the standard path-ordered product as
\begin{equation}
\hat{P}e^{\int d\hat\sigma \bM(\hat\sigma)} = P e^{\int d\sigma M(\sigma)},
\end{equation}
with
\begin{equation}
M(\sigma) = M_1(\sigma) - M_0(\sigma)^2.
\end{equation}
Here we should note that 
the rule for matrix multiplication in $M_0^2$ has extra sings as
\begin{equation}
\left(\begin{matrix} A & B \\ C & D \end{matrix}\right)
\left(\begin{matrix} A' & B' \\ C' & D' \end{matrix}\right)
=\left(\begin{matrix} AA'+(-1)^{c'} BC' & AB'+(-1)^{d'}BD' \\
DC' + (-1)^{a'} CA'  & DD' +(-1)^{b'} CB' \end{matrix}\right),
\end{equation}
where $a'$, $b'$, $c'$, $d'$ are $0$ $(1)$ when $A'$, $B'$, $C'$, $D'$ are
Grassmann even (odd), respectively.
In particular,
for $\bM$ given by (\ref{superM}), $M(\sigma)$ is written as \cite{Kraus2001}
\begin{equation}
M=\left(\begin{matrix}
-iA_\mu \del_\sigma X^\mu + \frac{i}{2} \Psi^\mu \Psi^\nu F_{\mu\nu} - T^2 &
i\Psi^\mu {\cal D}_\mu T \\
i\Psi^\mu {\cal D}_\mu T &
-iA_\mu \del_\sigma X^\mu + \frac{i}{2} \Psi^\mu \Psi^\nu F_{\mu\nu} -T^2
\end{matrix}\right),
\end{equation}
where $F_{\mu\nu}=\del_\mu A_\nu - \del_\nu A_\mu +i[A_\mu, A_\nu]$
is the field strength and ${\cal D}_\mu T = \del_\mu T + i[A_\mu, T]$ is
the covariant derivative of the tachyon field.
For more detail, see, eg, \cite{Asakawa2002}.

One of the most important feature is that the BSFT action
(\ref{BSFT action}) and the CS-term (\ref{CS term}) are invariant
under the gauge transformation,
\begin{equation}
\bM \to \bM' = -\hat{U}(\bX)^\dagger D\hat{U}(\bX) + \hat{U}(\bX)^\dagger \bM \hat{U}(\bX),
\label{gauge trans}
\end{equation}
with
\begin{equation}
\hat{U}(\bX)=\left(\begin{matrix} U(\bX) & 0 \\ 0 & U(\bX) \end{matrix}\right), \quad
U(x) \in U(N).
\end{equation}
For the tachyon and gauge fields, it turns out to be
\begin{equation}
\begin{split}
A_\mu(x) &\to A'_\mu(x) = U(x)^\dagger A_\mu(x) U(x) + i U(x)^\dagger \del_\mu U(x), \\
T(x) &\to  T'(x) = U(x)^\dagger T(x) U(x),
\end{split}
\end{equation}
which is nothing but the gauge transformations of the gauge field and the
tachyon field in the adjoint representation.

\subsection{Boundary interaction for spherical D2-brane}
\label{app:boundary interaction}

For the tachyon profile (\ref{profile}) and $A_\mu=0$,
the matrix (\ref{superM}) can be written as
\begin{equation}
\bM=\left(\begin{matrix}
0 & u(\sigma_i \bX^i - R) \\
u(\sigma_i \bX^i - R) & 0
\end{matrix}\right).
\end{equation}
Performing the gauge transformation \eqref{gauge trans} with 
$U=U_{(N/S)}$ in \eqref{unitary matrix}, 
$\bM$ transforms into $\bM'$ as (\ref{gauge trans}),
which is explicitly written as
\begin{align}
	\bM'&=u{\mathbf B} + \bA \nonumber \\
&\equiv  u\,
\sigma_1 \otimes \lambda(\bX)
+
 {\bf 1}_2
 \otimes
 (-U(\bX)^\dagger DU(\bX)),
 \label{PandA}
\end{align}
with
\begin{equation}
\lambda(\bX) 
=
\left(\begin{matrix}
|\bX|-R & 0 \\ 0 & -|\bX|-R
\end{matrix}\right) 
\equiv
\left(\begin{matrix}
t(\bX) & 0 \\ 0 & s(\bX)
\end{matrix}\right).
\label{diagonal lambda}
\end{equation}
Here we note that the (super) path-order product
$\hat{P}e^{\int d\hsigma \bM'(\hat\sigma)}$
appearing in (\ref{super path order})
can be expanded as
\begin{align}
	\hat{P}&e^{\int d\hsigma(u{\mathbf B}+\bA)}
	= \hat{P}e^{u \int d\hsigma {\mathbf B}} \nonumber \\
&+ \sum_{n=1}^\infty \int d\hsigma_1\cdots d\hsigma_n
	\hat{T}_{{\mathbf B}}(\hat{2\pi},\hsigma_n) \bA(\hsigma_n) \hat{T}_{{\mathbf B}}(\hsigma_n,\hsigma_{n-1}) \cdots
	\hat{T}_{{\mathbf B}}(\hsigma_2,\hsigma_1) \bA(\hsigma_n) \hat{T}_{{\mathbf B}}(\hsigma_1,\hat{0}),
\label{boundary action2}
\end{align}
where
\begin{equation}
	\hat{T}_{{\mathbf B}}(\hsigma_f,\hsigma_i) \equiv
	\hat{P} e^{u \int_{\hsigma_i}^{\hsigma_f} d\hat{\tau}  {\mathbf B}(\hat\tau)}.
\end{equation}
The ``transfer matrix'' $\hat{T}_{{\mathbf B}}(\hsigma_f,\hsigma_i)$ can also be expanded as
\begin{align}
	\hat{T}_{{\mathbf B}}&(\hsigma_f,\hsigma_i) = T_{-{B}_0^2}(\sigma_f,\sigma_i)
\nonumber \\
&+\sum_{n=1}^\infty u^n \int d\sigma_1 \cdots d\sigma_n \Bigl(
T_{-{B}_0^2}(\sigma_f,\sigma_n) {B}_1(\sigma_n) T_{-{B}_0^2}(\sigma_n,\sigma_{n-1})
\nonumber \\
&\hspace{5cm} \cdots
T_{-{B}_0^2}(\sigma_2,\sigma_1) {B}_1(\sigma_1) T_{-{B}_0^2}(\sigma_1,\sigma_{i})
\Bigr),
\label{T-hat expand}
\end{align}
where
\begin{equation}
	{\mathbf B}(\hsigma) \equiv {B}_0(\sigma)+\theta {B}_1(\sigma),
\end{equation}
and
\begin{equation}
	{T}_{-{B}_0^2}(\sigma_f,\sigma_i) \equiv
	P e^{-u^2\int_{\sigma_i}^{\sigma_f} d{\tau}  {B}_0(\tau)^2}.
\end{equation}
Since $\lambda(x)$ is diagonal,
$T_{-{B}_0^2}(\sigma_f, \sigma_i)$ and ${B}_1(\sigma)$
can be written as
\begin{align}
	T_{-{B}_0^2}(\sigma_f, \sigma_i)&=
{\bf 1}_2 \otimes
\left( \begin{matrix}
Pe^{-u^2\int_{\sigma_i}^{\sigma_f} d\tau t(X(\tau))^2 }  & 0 \\
0 & Pe^{-u^2\int_{\sigma_i}^{\sigma_f} d\tau s(X(\tau))^2 }
\end{matrix} \right),  \\
{B}_1(\sigma) &=
\sigma_1 \otimes
\left( \begin{matrix}
u \psi^\mu \del_\mu t(X) & 0 \\
0 & u \psi^\mu \del_\mu s(X)
\end{matrix} \right)  .
\end{align}
The element
$Pe^{-u^2\int_{\sigma_i}^{\sigma_f} d\tau s(X(\tau))^2}$
vanishes in the limit of $u\to\infty$
because $s(x)\ne 0$ for ${}^\forall x\in \R^3$.
Therefore, if we are interested only in this limit, we can regard
$T_{-{B}_0^2}(\sigma_f, \sigma_i)$ as
\begin{align}
	T_{-{B}_0^2}(\sigma_f, \sigma_i)&\underset{u\to\infty}{\sim}
 {\bf 1}_2  \otimes
Pe^{-u^2\int_{\sigma_i}^{\sigma_f} d\tau t(X(\tau))^2} \,
P_0,
\end{align}
where $P_0\equiv \left(\begin{matrix} 1 & 0\\ 0 & 0 \end{matrix} \right)$ is
the projection operator to a one-dimensional subspace of
the two-dimensional Chan-Paton space
and the symbol
"$\underset{u\to\infty}{\sim}$" used here
expresses that the both hand sides
are identical in the limit of $u\to\infty$.

Since the matrices ${B}_1$ in (\ref{T-hat expand})
are sandwiched by the operators $T_{-{B}_0^2}$, which is proportional
to the projection operator in the limit of $u\to\infty$,
${B}_1(\sigma)$ can be regarded as
\begin{align}
	{B}_1(\sigma)&\underset{u\to\infty}{\sim}
 \sigma_1 \otimes
\psi^\mu(\sigma)\del_\mu t(X(\sigma))\,
P_0,
\label{P1}
\end{align}
and thus the transfer operator $\hat{T}_{{\mathbf B}}(\hat\sigma_f,\hat\sigma_i)$
in the expansion (\ref{boundary action2})
can also be regarded to be proportional to the projection operator $P_0$.
This means that the tachyon profile after the gauge transformation
is equivalent to
\begin{equation}
T(\bX)  \underset{u\to\infty}{\sim}
u\, \left(\begin{matrix}  t(\bX) & 0 \\
0 & 1
\end{matrix}\right)
= u\, \left(\begin{matrix} \sqrt{ \sum_{i=1}^3 \bX_i^2} - R & 0 \\
0 & 1
\end{matrix}\right) ,
\label{tachyon profile}
\end{equation}
in the limit of $u\to\infty$
and the gauge connection can be regarded as the projection
of the pure gauge connection $U^\dagger DU$
to the one-dimensional subspace of the two-dimensional Chan-Paton space
by the projection operator $P_0$,
\begin{equation}
A(x) \underset{u\to\infty}{\sim}   P_0 \left( -i U(x)^\dagger d U(x) \right) P_0,
\label{gauge field2}
\end{equation}
{which gives a non-trivial gauge connection (\ref{monopole potential}).}

\subsection{D2-brane tension}
\label{sec:D-brane interpretation}

The tension is evaluated also from the BSFT action  (\ref{BSFT action}).
To this end, we ignore the gauge potential (\ref{gauge field2}) 
and only consider the tachyon profile (\ref{tachyon profile})
for simplicity.

Let us first separate $X^i(\sigma)$ ($i=1,2,3$) into
the zero-mode $x^i$ and the nonzero-modes $\hat{X}^i(\sigma)$
and expand
$t(X(\sigma))=\sqrt{\sum_{i=1}^3 X_i(\sigma)^2} - R$ as
\begin{align}
t(X(\sigma))&=t(x) + \sum_{i=1}^3 \del_i t(x) \hat{X}^i(\sigma) + {\cal O}(\hat{X}(\sigma)^2) \nonumber \\
&= r - R + \sum_{i=1}^3  \frac{x_i}{r} \hat{X}^i(\sigma) + {\cal O}(\hat{X}(\sigma)^2).
\label{expand}
\end{align}
In performing the functional integral over $\hat{X}^i(\sigma)$,
the contribution from
the ${\cal O}(\hat{X}(\sigma)^2)$ terms can be regarded as
perturbations whose coupling constants are of the order of ${\cal O}(u^{-1})$,
which vanishes in the limit of $u\to\infty$.
Therefore, by repeating the computation in \cite{Kraus2001, Takayanagi2000},
the BSFT action can be evaluated as
\begin{align}
S(T)&=\sqrt{2} T_{3}\int d^{4}x e^{-2\pi u^2 t(x)^2 }
\biggl( F(4\pi u^2 ( \del_i t(x) )^2 )  + {\cal O}(u^{-1}) \biggr) \nonumber \\
&=2\pi T_{3}\int d^{4}x
\frac{u}{\sqrt{\pi}}
e^{-2\pi u^2 (r-R)^2 }
\bigl(1 + {\cal O}(u^{-1}) \bigr),
\end{align}
where the function
$F(y)\equiv \frac{4^y y \Gamma(y)^2}{2\Gamma(2y)} \sim \sqrt{\pi y}$
($y \gg 1$) comes from the linear term
in the expansion (\ref{expand}).
As shown in the appendix \ref{app:radial delta},
the $u\to\infty$ limit of the leading term of the $u$-expansion
of the integrand is
a radial delta-function,
\begin{equation}
\lim_{\mu\to\infty} \frac{u}{\sqrt{\pi}}
e^{-2\pi u^2 (r-R)^2 } = \delta(r-R),
\end{equation}
with the property \eqref{delta def}.
Recalling the tensions of D$p$-brane and D$(p-1)$-brane are
related as $2\pi T_{p}=T_{p-1}$, we see
\begin{equation}
S(T)  \underset{u\to\infty}{\to}  T_{2} {\rm vol}(S^2_R) \int dt
\end{equation}
where ${\rm vol}(S^2_R)$ is the volume of the 2-sphere
with the radius $R$.
It shows that the tension of this spherical object is that of the D2-brane.

We can repeat the same analysis for the RR sector. 
However, since the contributions from the non-zero modes of the 
bosons $X^M(\sigma)$ and the fermions $\Psi^M(\sigma)$ cancel with 
other, we have only to consider the integration over the zero modes in
the evaluation \cite{Kraus2001, Takayanagi2000}. 
Therefore the analysis is identical to that we have done in 
the section \ref{subsec:direct}.


\providecommand{\href}[2]{#2}\begingroup\raggedright\endgroup

\end{document}